\documentclass[letterpaper,rmp, natbib]{revtex4-1}

\usepackage{url}
\usepackage{graphicx}
\usepackage{times}
\usepackage{multirow}

\usepackage{url}  % Formatting web addresses  
\usepackage{ifthen}  % Conditional 
\usepackage[utf8]{inputenc} %unicode support

\begin{document}

\title{Partisan Asymmetries in Online Political Activity}

\author{Michael D. Conover$^1$}
\author{Bruno Gon\c calves$^2$}\email{b.goncalves@neu.edu} %
\author{Alessandro Flammini$^1$}
\author{Filippo Menczer$^1$}

\affiliation{${}^{1}$Center for Complex Networks and Systems Research,\\School of Informatics and Computing\\Indiana University\\
\\${}^{2}$College of Computer and Information Sciences\\Northeastern University}

%%%%%%%%%%%%%%%%%%%%%%%%%%%%%%%%%%%%%%%%%%%%%%
%%                                          												   %%
%% Enter the authors' addresses here        								 	   %%
%%                                          											            %%
%%%%%%%%%%%%%%%%%%%%%%%%%%%%%%%%%%%%%%%%%%%%%%

\begin{abstract}
We examine partisan differences in the behavior, communication patterns and social interactions of more than $18,000$ politically-active Twitter users to produce evidence that points to changing levels of partisan engagement with the American online political landscape. Analysis of a network defined by the communication activity of these users in proximity to the $2010$ midterm congressional elections reveals a highly segregated, well clustered partisan community structure. Using cluster membership as a high-fidelity ($87\%$ accuracy) proxy for political affiliation, we characterize a wide range of differences in the behavior, communication and social connectivity of left- and right-leaning Twitter users.  We find that in contrast to the online political dynamics of the $2008$ campaign, right-leaning Twitter users exhibit greater levels of political activity, a more tightly interconnected social structure, and a communication network topology that facilitates the rapid and broad dissemination of political information.  
\end{abstract}

\maketitle

\section{Introduction}

Digitally-mediated communication has become an integral part of the American political landscape, providing citizens access to an unprecedented wealth of information and organizational resources for political activity.  So pervasive is the influence of digital communication on the political process that almost one quarter ($24\%$) of American adults got the majority of their news about the 2010 midterm congressional elections from online sources, a figure that has increased three-fold since the Pew Research Center began monitoring the statistic during the $2002$ campaign~\cite{pew_campaign2010}. Relax the constraint that a majority of a person's political news and information must come from online sources and the figure jumps to include the $54\%$ of adult Americans who went online in $2010$ to get political information. Critically, this activity precipitates tangible changes in the beliefs and behaviors of voters, with $35\%$ of internet users who voted in $2010$ reporting that political information they saw or read online made them decide to vote for or against a particular candidate~\cite{pew_campaign2010}.

Within this ecosystem of digital information resources, social media platforms play an especially important role in facilitating the spread of information by connecting and giving voice to the voting public~\cite{bennett2003new,aday10-1,farrell08}. Networked and unmoderated, social media are characterized by the large-scale creation and exchange of user-generated content~\cite{kaplan2010users}, a production and consumption model that stands in stark contrast to the centralized editorial and distribution processes typical of traditional media outlets~\cite{Benkler:2006kx,sunstein_republic20}.

In terms of political organization and engagement, the benefits of social media use are many. For voters, social media make it easier to share political information, draw attention to ideological issues, and facilitate the formation of advocacy groups with low barriers to entry and participation~\cite{tolbert2003unraveling,garrett2006protest}. The ease with which individual voters can connect with one another directly also makes it easier to aggregate small-scale acts, as in the case of online petitions, fundraising, or web-based phonebanking~\cite{land2009networked}. Together, these features contribute to the widespread use of social media for political purposes among the voting public, with as many as $21\%$ of online adults using social networking sites to engage with the $2010$  congressional midterm elections~\cite{pew_socmed2010}. Moreover, a survey by the Pew Internet and American Life Project finds that online political activity is correlated with more traditional forms of political participation, with individuals who use blogs or social networking sites as a vehicle for civic engagement being more likely to join a political or civic group, compared to other internet users~\cite{pew_internetcivic}.  

Likewise, candidates and traditional political organizations benefit from a constituency that is actively engaged with social media, finding it easier to raise money, organize volunteers and communicate directly with voters who use social media platforms~\cite{edelmen-09}.  Social media also facilitate the rapid dissemination of political frames, making it easy for key talking points to be communicated directly to a large number of constituents, rather than having to subject messages to the traditional media filter. 

Considered in this light, it becomes clear why social media were argued to have played such an important role in the political success of the Democratic party in the 2008 presidential and congressional elections~\cite{nyt_obama08-1,adage_obama08-1,busweek_obama08-1}.  Survey data from the Pew Research Center showed that, along the seven dimensions used to measure online political activity, Obama voters were substantially more likely to use the internet as an outlet for political activity~\cite{pew_campaign2008}.  In particular, Obama voters were more likely than McCain voters to create and share political cselectontent, and to engage politically on an online social network~\cite{pew_campaign2008}. Moreover, a 2009 Edelman report found that in addition to a thirteen million member e-mail list, the Obama campaign enjoyed twice as much web traffic, had four times as many YouTube viewers and five times more Facebook friends compared to the McCain campaign~\cite{edelmen-09}. While the direct effect of any one media strategy on the success of a campaign is difficult to assess and quantify, the data show that Obama campaign had a clear advantage in terms of online voter engagement. 

Motivated by the connection between the widely reported advantage in on-line mobilization and the result of the $2008$ presidential election, we seek to understand structural shifts in the American political landscape with respect to partisan asymmetries in online political engagement. We work toward this goal by examining partisan differences in the behavior, communication patterns and social interactions of more than $18,000$ politically-active users of the popular social media platform Twitter. 
% R&R 2.3.1
Among all social media services, Twitter makes an appealing analytical target for a number of reasons: the public nature of its content, the accessibility of the data through APIs, a strong focus on news and information sharing, and its prominence as a platform for political discourse in America and abroad~\cite{kwak2010twitter,howard2011opening}.  These features make a compelling case for using this platform to study partisan political activity.

For this analysis we build on the findings of a previous study which established the macroscopic structure of domestic political communication on Twitter, a social networking platform that allows individuals to create and share brief 140-character messages. In that work we employed clustering techniques and qualitative content analysis to demonstrate that the network of political retweets exhibits a highly segregated, partisan structure~\cite{conover11-1}. Despite this segregation, we found that politically left- and right-leaning individuals engage in interaction across the partisan divide using mentions, a behavior strongly correlated with a type of cross-ideological provocation we term `content injection.'  

Having established the large-scale structure of these communication networks, in this study we employ a variety of methods to provide a more detailed picture of domestic political communication on Twitter.  We characterize a wide range of differences in the behavior, communication, geography and social connectivity of thousands of politically left- and right-leaning users. Specifically, we demonstrate that right-leaning Twitter users exhibit greater levels of political activity, tighter social bonds, and a communication network topology that facilitates the rapid and broad dissemination of political information, a finding that stands in stark contrast to the online political dynamics of the 2008 campaign.  

With respect to individual-level behaviors, we find that right-leaning Twitter users produce more than 50\% more total political content and devote a greater proportion of their time to political discourse.  Right-leaning users are also more likely to use hyperlinks to share and refer to external content, and are almost twice as likely than left-leaning users to self-identify their political alignment in their profile biographies.  At the individual level, these behavioral factors paint a picture of a right-leaning constituency comprised of highly-active, politically-engaged social media users, a trend we see reflected in the communication and social networks in which these individuals participate.  

Regarding connectivity patterns among users in these two communities we report findings related to three different networks, described by the set of explicitly declared follower/followee relationships, mentions, and retweets. Casting the declared follower network as the social substrate over which political information is most likely to spread, we find that right-leaning users exhibit a greater propensity for mutually-affirmed social ties, and that right-leaning users tend to form connections with a greater number of individuals in total compared to those on the left. With respect to the way in which information actually propagates over this substrate in the form of retweets, right-leaning users enjoy a network structure that is more likely to facilitate the rapid and broad dissemination of political information. Additionally, right-leaning users exhibit a higher probability to rebroadcast content from and to be rebroadcast by a large number of users, and are more likely to be members of high-order retweet network $k$-cores and k-cliques, structural features that are associated with the efficient spreading of information and adoption of political behavior and opinions. Pointing definitively to a vocal, socially engaged, densely interconnected constituency of right-leaning users, these topological and behavioral features provide a significantly more nuanced perspective on political communication on this important social media platform.  Moreover, through its use digital trace data to illuminate a complex sociological phenomenon, this article illustrates the explanatory power of data science techniques and underscores the potential of this burgeoning scientific epistemology.  

\section{Platform \& Data}

\subsection{The Twitter Platform}
\label{sec:platform}
Twitter is a popular social networking and microblogging site where users can post 140-character messages containing text and hyperlinks, called \emph{tweets}, and interact with one another in a variety of ways. In the present section we describe four of the platform's key features: follow relationships, retweets, mentions, and hashtags.

Twitter allows each user to broadcast tweets to an audience of users who have elected to subscribe to the stream of content he or she produces. The act of subscribing to a user's tweets is known as \emph{following}, and represents a directed, non-reciprocal social link between two users.  From a content consumption perspective, each user can sample tweets from a variety of content streams, including the stream of tweets produced by the users he or she follows, as well as the set of tweets containing specific keywords known as hashtags.  

Hashtags are tokens prepended with a pound sign (\emph{i.e.} {\tt \#token}) which, when displayed, function as a hyperlink  to the stream of recent tweets containing the specified tag~\cite{Java:2007lk}. While they can be used to specify the topic of a tweet (\emph{i.e.} {\tt \#oil} or {\tt \#taxes}), when used in political communication hashtags are commonly employed to identify one or more intended audiences, as in the case of the most popular political hashtags,  {\tt \#tcot} and {\tt \#p2}, acronyms for ``Top Conservative on Twitter'' and ``Progressives 2.0,'' respectively. In this way, hashtags function to broaden the audience of a tweet, extending its visibility beyond a person's immediate followers to include all users who seek out content associated with the tag's topic or audience.  For this reason, as outlined in Section~\ref{sec:discovering_political_communication}, we restrict our analysis to the set of tweets containing political hashtags, ensuring that the content under study is broadly public and expressly political in nature.

% R&R : 2.1 (Replies v. Mentions)
In addition to broadcasting tweets to the public at large, Twitter users can interact directly with one another in two primary ways: retweets and mentions. Retweets often act as a form of endorsement, allowing individuals to rebroadcast content generated by other users, thus raising the content's visibility~\cite{boyd08}. Mentions allow someone to address a specific user directly through the public feed, or, to a lesser extent, refer to an individual in the third person. In this study, we differentiate between mentions that occur in the body of the tweet and those that occur at the beginning of a tweet, as they correspond to distinct modes of interaction.  Mentions located at the beginning of a tweet are known as `replies', and typically represent actual engagement, while mentions in the body of a tweet typically constitute a third-person reference~\cite{honeycutt08-1}. Together, retweets and mentions act as the primary mechanisms for explicit, public user-user interaction on Twitter. 

\subsection{Data}
\label{sec:data}

The analysis described in this article relies on data collected from the Twitter `gardenhose' streaming API~\footnote{\url{dev.twitter.com/pages/streaming_api/}} between September ${1}^{st}$ and January $7^{th}$, 2011 --- the eighteen week period surrounding the November $4^{th}$ United States congressional midterm elections. The gardenhose provides a sample of approximately 10\% of the entire Twitter corpus in a machine-readable format. Each tweet entry is composed of several fields, including a unique identifier, the content of the tweet (including hashtags and hyperlinks), the time it was produced, the username of the account that produced the tweet, and in the case of retweets or mentions, the account names of the other users associated with the tweet. 

From this eighteen week period we collected data on 6,747 right-leaning users and 10,741 left-leaning users, responsible for producing a total of 1,390,528 and 2,420,370 tweets, respectively. It's useful to note that we evaluate all gardenhose tweets associated with each user, rather than just those containing political hashtags, in order to facilitate comparisons between the two groups in terms of relative proportions of attention allocated to political communication.  

\section{Methodology}
\label{sec:methodology}

In order to examine differences in the behavior and connectivity of left- and right-leaning Twitter users we rely on the political hashtags and partisan cluster membership labels established in a previous study on political polarization.  In addition to reviewing the approach used to establish these features, we %provide data in support of the claim 
show that the networks and communities under study are representative of domestic political communication on Twitter in general.  

\subsection{Identifying Political Content}
\label{sec:discovering_political_communication}

As outlined in Section~\ref{sec:platform}, hashtags are used to specify the topic or intended audience of a tweet, and allow a user to engage a much larger potential audience than just his or her immediate followers.  We define the set of pertinent political communication as any tweet containing at least one political hashtag. While an individual can engage in political communication without including a hashtag, the potential audience for such content is limited primarily to his or her immediate followers.  Moreover, restricting our analysis to tweets which have been expressly identified as political in nature allows us to define a high-fidelity corpus, avoiding the risk of introducing undue noise through the use of topic detection strategies~\cite{landauer1998introduction,blei2003latent}. 

%%%%%%%%%
%
% TABLE: Political Hashtags
%
\begin{table}
\caption{\label{table:political_tags}Political hashtags related to {\tt \#p2} and {\tt \#tcot} (acronyms for `Progressives 2.0' and `Top Conservatives on Twitter'). Tweets containing any of these were included in our sample.}
\begin{tabular}{lp{0.69\columnwidth}}
\hline
Just \texttt{\#p2} & 
\scriptsize
\texttt{\#casen \#dadt \#dc10210 \#democrats \#du1 \#fem2 \#gotv \#kysen \#lgf \#ofa \#onenation \#p2b \#pledge \#rebelleft \#truthout \#vote \#vote2010 \#whyimvotingdemocrat \#youcut} \\
\hline
Both & 
\scriptsize
\texttt{\#cspj \#dem \#dems \#desen \#gop \#hcr \#nvsen \#obama \#ocra \#p2 \#p21 \#phnm \#politics \#sgp \#tcot \#teaparty \#tlot \#topprog \#tpp \#twisters \#votedem } \\
\hline
Just \texttt{\#tcot} &
\scriptsize
\texttt{\#912 \#ampat \#ftrs \#glennbeck \#hhrs \#iamthemob \#ma04 \#mapoli \#palin \#palin12 \#spwbt \#tsot \#tweetcongress \#ucot \#wethepeople } \\
\hline
\end{tabular}
\end{table}

%%%%%%%%%

To isolate a representative set of political hashtags and to avoid introducing bias into the dataset we performed a simple algorithmic hashtag discovery procedure. We began by seeding our sample with the two most popular political hashtags, {\tt \#p2} (``Progressives 2.0'') and {\tt \#tcot} (``Top Conservatives on Twitter''). For each seed we identified the set of hashtags with which it co-occurred in at least one tweet, and ranked the results using the Jaccard coefficient. For a set of tweets $S$ containing a seed hashtag, and a set of tweets $T$ containing a second hashtag, the Jaccard coefficient between $S$ and $T$ is
\begin{equation}
\sigma(S, T) = \frac{\left| S \cap T \right|} {\left| S \cup T\right|}.
\end{equation}
Thus, when the tweets in which both seed and the second hashtag occur make up a large portion of the tweets in which either occurs, the two are deemed to be related. Using a similarity threshold of 0.005 we identified sixty six unique hashtags (Table~\ref{table:political_tags}), eleven of which were excluded due to overly-broad or ambiguous meanings (Table~\ref{table:excluded_tags}).  While it is a common practice among spammers to contribute content to popular hashtag streams, we do not believe this phenomenon plays a substantial role in a shaping the structure of the sample data.  During a previous study we found that of 1,000 manually-inspected accounts identified by this methodology fewer than 3\% corresponded to foreign language or spam activity~\cite{conover11-1}.  

\begin{table}
\caption{Hashtags excluded from the analysis due to ambiguous or overly broad meaning.}
\begin{tabular}{lp{0.58\columnwidth}}
\hline
Excl. from \texttt{\#p2} & 
\scriptsize \texttt{\#economy \#gay \#glbt \#us \#wc \#lgbt} \\
\hline
Excl. from both & 
\scriptsize \texttt{\#israel \#rs} \\
\hline
Excl. from \texttt{\#tcot} &
\scriptsize \texttt{\#news \#qsn \#politicalhumor} \\
\hline
\end{tabular}
\label{table:excluded_tags}
\end{table}

\subsection{Representativeness}
\label{sec:Representativeness}

\begin{figure}
\begin{center}
%\centerline{
	\includegraphics[width=\textwidth,]{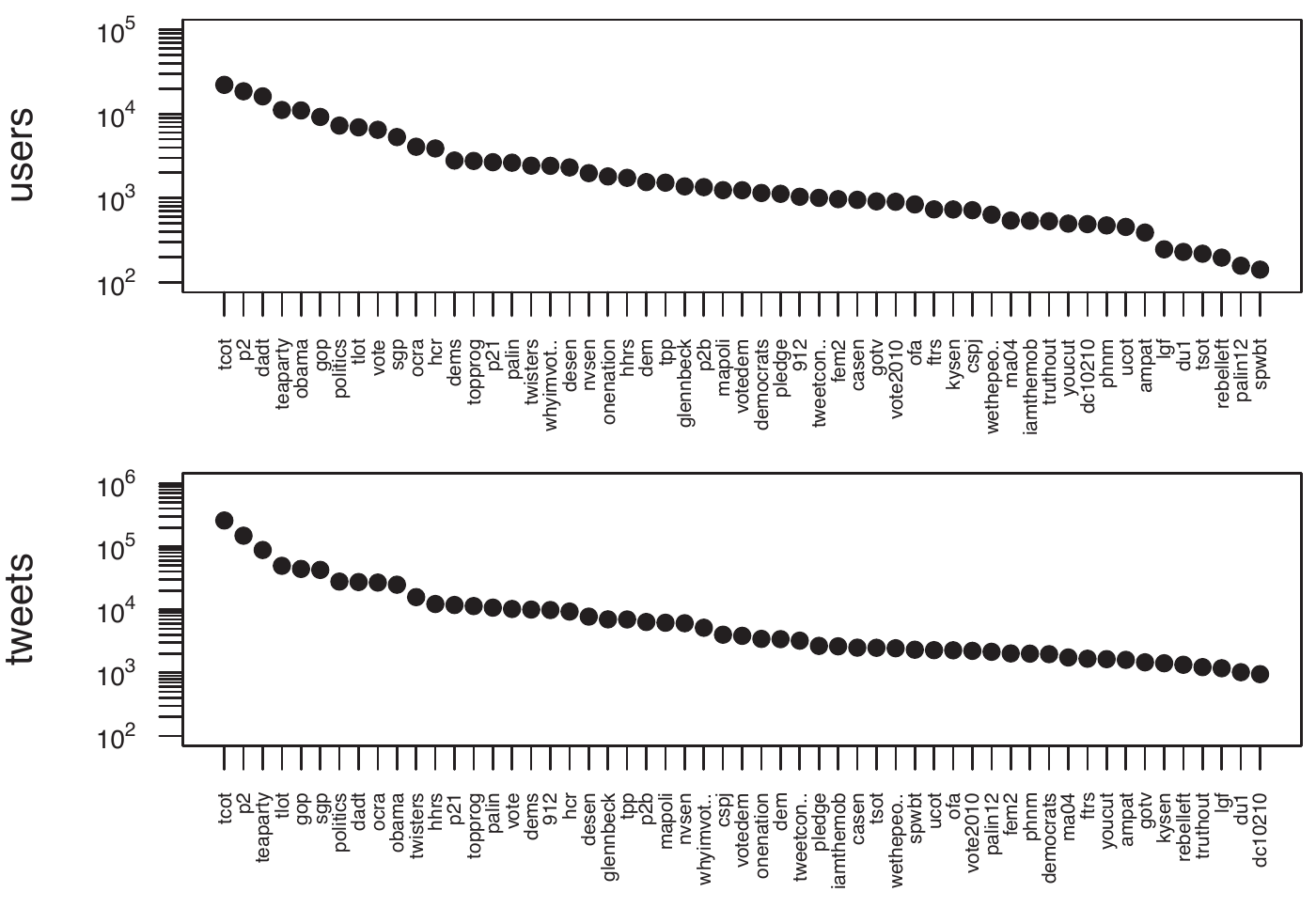}
%}
\vspace{-1em}
\caption{Hashtag popularity decay in terms of total number of tweets and users associated with each tag. On the horizontal axis tags have been ordered according to one of the two popularity measures: number of tweets (bottom) and users (top). The roughly exponential decay indicates that the inclusion of additional hashtags is unlikely to result in a substantial increase in the size of the corpus.}
\label{fig:hashtag_popularity}
\end{center}
\end{figure}

Using the technique outlined above we identified many high-profile political hashtags, and with them the majority of tweets and users associated with domestic political communication on Twitter.  Supporting this claim, Figure~\ref{fig:hashtag_popularity} shows a roughly exponential decay in hashtag popularity as measured in terms of number of users or tweets associated with the hashtag. This sharp decay in the tag popularity indicates that the inclusion of additional political hashtags is not likely to substantially increase the size or alter the structure of the corpus.

%%%%%% 
%
%  FIGURE:  Hashtag Popularity (Users & Tweets)
% 
%
%%%%%%%

\begin{figure}
\begin{center}
	\includegraphics[width=\textwidth,]{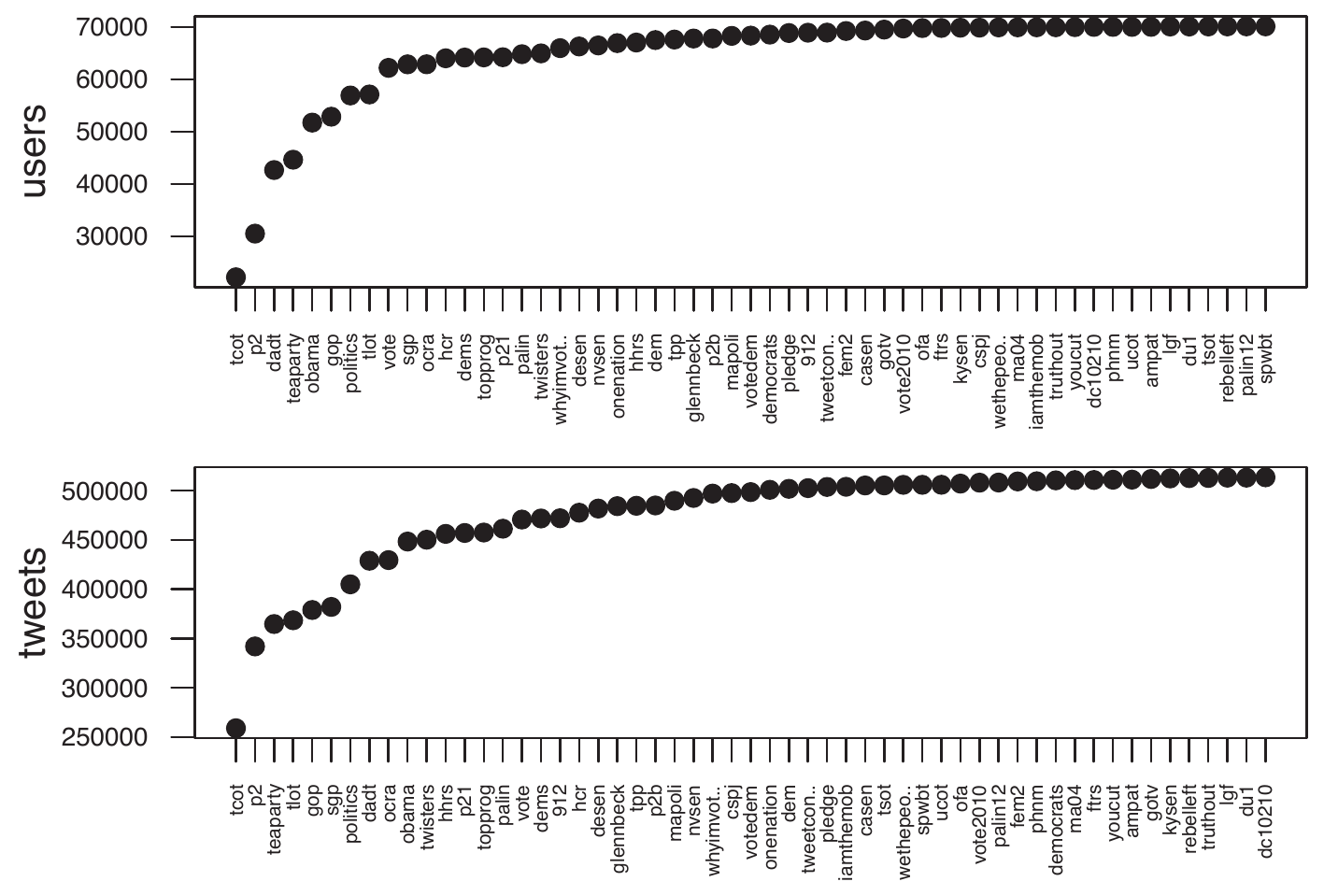}
\caption{Size of the set of unique users and tweets resulting from the inclusion of additional hashtags. Axes are ordered according to the total number of tweets (top) and users (bottom) associated with each tag.}
\label{fig:aggregate_tweets_users}
\end{center}
\end{figure}

This claim is also supported by Figure~\ref{fig:aggregate_tweets_users}, which shows that there is a strong effect of diminishing returns with respect to the observed number of unique users and tweets as the number of hashtags included in our analysis increases.  This effect is due to the fact that many tweets are annotated with multiple hashtags, and many users utilize several different hashtags over the course of the study period.  As a result, the inclusion of a single hashtag may result in the inclusion of many tweets and users also redundantly associated with other hashtags. 

%%%%%% 
%
%  FIGURE: Aggregate Unique Tweets & Users 
% 
%
%%%%%%%

To further support the claim that sampling based on this set of hashtags produces a representative set of political tweets, we selected all the tweets in the gardenhose from the study period that included any one of 2500 hand-selected political keywords related to the 2010 elections~\cite{ratkiewicz11-1}.   We considered only the 312,560 tweets in this set containing a hashtag because we use this characteristic to define public political communication on Twitter. We found that 26.4\% of these tweets are covered by our target set of hashtags. Furthermore, among the ten most popular hashtags not included in our target set (\texttt{\#2010memories}, \texttt{\#2010disappointments}, \texttt{\#ff}, \texttt{\#p2000}, \texttt{\#2010}, \texttt{\#business}, \texttt{\#uk}, \texttt{\#newsjp}, \texttt{\#asia}, \texttt{\#sports}), only one is explicitly political and its volume accounts for less than 2\% of public political communication.  This coverage confirms that we have isolated a substantial and representative sample of political communication on Twitter.  

\subsection{Inferring Political Identities from Communication Networks}
\label{sec:networks_of_political_communication}
In a previous study we used the set of political tweets from the six weeks preceding the $2010$ midterm election to build a network representing political retweet interactions among Twitter users. In this network an edge runs from a node representing user $A$ to a node representing user $B$ if $B$ retweets content originally broadcast by $A$, indicating that information has propagated from $A$ to $B$.  This network consists of $23,766$ non-isolate nodes among a total of $45,365$, with $18,470$ nodes in its largest connected component and $102$ nodes in the next-largest component.  We describe the construction of an analogous network of political mentions in Section~\ref{sec:mention_network}.

% R&R 2.4.1
Using a combination of network clustering algorithms and manually-annotated data we determined that the network of political retweets neatly divides the population of users in the largest connected component into two distinct communities~(Figure~\ref{fig:network_layout})~\cite{conover11-1}.  In brief, we used Rhaghavan's label propagation method seeded with node labels determined by Newman's leading eigenvector modularity maximization method to assign cluster membership to each node~\cite{Newman:2006zr,Raghavan:2007fk}.  The final community assignments are consistent and robust to fluctuations in starting conditions~\cite{conover11-1}. To determine whether these communities were composed of users from the political left and right, respectively, we used qualitative content analysis evaluate the tweets produced by 1,000 random users appearing in the intersection of the mention and retweet networks~\cite{krippendorf04,kolbe91}.  

% R&R 2.5
To establish the reproducibility of these results we had two authors, working independently, determine whether the content of a user's tweets express a `left', `right' or `undecidable' political identity according to the coding rubric developed in a previous study~\cite{conover11-1}.  These annotations were compared against the work of an independent non-author judge, and using a well-established measure of inter-annotator agreement we report `nearly perfect' inter-annotator agreement between author and non-author annotations for the `left' and `right' classes (Cohen's Kappa values of .80 and .82, respectively) and `fair to moderate' agreement for the `undecidable' category (Cohen's Kappa value of .42)~\cite{krippendorf04,kolbe91}. From these high levels of inter-annotator agreement we conclude that an objective outside party would be able to reproduce our class assignments for most users.

Based on this content analysis, we determined that the retweet network communities are highly politically homogeneous, consisting of $80.1\%$ left- and $93.4\%$ right-leaning users, respectively (Table~\ref{tab:network_clusters})~\cite{conover11-1}.  In this study we use network community membership as a proxy for the political identities of all $18,470$ users in the largest connected component of the retweet network, and hereafter focus on the behavior of these users.  Based on the relative proportions of right- and left-leaning users identified during the qualitative content analysis stage, this mechanism results in correct predictions for $87.3\%$ of users in the largest connected component of the retweet network~\cite{socialcom_prediction}.  

%%%%%%
%
% TABLE:  ICWSM Retweet Community Stats
\begin{table*}
\caption{\label{tab:network_clusters}Partisan composition of retweet cluster communities as determined through manual annotation of 1,000 random users. (See \S~\ref{sec:networks_of_political_communication}).}
\centerline{
	\begin{tabular}{lcccccc}
	\hline
	 & Cluster & Left & Right & Undecidable & \# Nodes \\
	\hline
	\multirow{2}{*} 	& A (Top) & $1.19\%$ & $93.4\%$ & $5.36\%$ & $7,115$ \\
				& B (Bottom) & $80.1\%$ & $8.71\%$ & $11.1\%$ & $11,355$ \\
		\hline
	\end{tabular}
}
\vspace{-1em}
\end{table*}
%
%%%%%%%

In the following sections we leverage these data to explore, in detail, how users from the political left and right utilize this important social media platform for political activity in different ways.

\section{Behavior: Individual-level Political Activity}
Before examining structural differences in the social and communication networks of left- and right-leaning Twitter users, we first focus on political activity at the individual level. In this section we compare users in the left- and right-leaning communities in terms of their relative rates of content production, the amount of attention they allot to political communication, their respective rates of political self-identification, and their propensity for sharing information resources in the form of hyperlinks.

% R&R 1.1.4
Right-leaning users are substantially more active and politically engaged with this social media platform.  Specifically, our analysis shows that left-leaning users produce less total political content, allocate proportionally less time to creating political content, are less likely to reveal their political ideology in their profile biography, and are less likely to share resources in the form of hyperlinks.  All of these findings stand in stark contrast to survey data and media reportage of the 2008 online political dynamics, and provide evidence in support of the notion that right-leaning voters are becoming more politically engaged online.

\subsection{Political Communication}
\label{sec:political_communication}
\begin{figure}
\begin{center}
\centerline{
\includegraphics[width=\textwidth,]{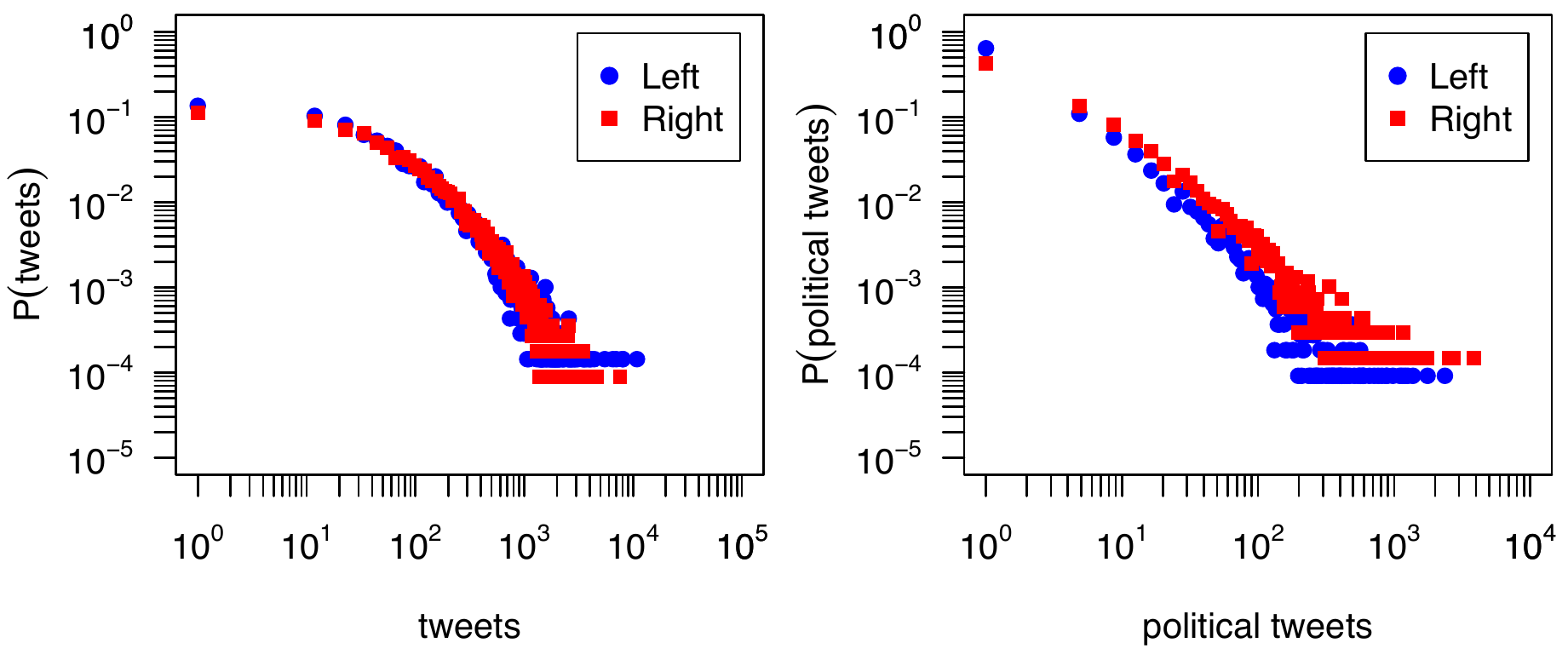}
}
\vspace{-1em}
\caption{Total number of tweets produced by right- and left-leaning users (left) compared to the total number of political tweets produced by users in each group.  While both groups produce a comparable amount of content in general, right-leaning users produce a much larger number of political tweets despite comprising fewer users in total. We observe that users' behavior tends to be broadly distributed, with many individuals creating relatively few tweets, while a few individuals produce substantially larger volumes of content. Note, however, that this sample includes only users who produced at least one political hashtag, rather than a random sample among all Twitter users, a feature likely responsible for the low number of users who produce few total tweets.}
\label{fig:ab_hist_tweets}
\end{center}
\end{figure}

From the perspective of leveraging social media for political organization, the baseline level of activity among a constituency is one of the most important characteristics of a population. Figure~\ref{fig:ab_hist_tweets} shows that while left- and right-leaning users produce approximately the same number of tweets per user, right-leaning individuals actually produce $54\%$ more total political content despite comprising fewer users altogether.  This trend is the result of divergent priorities among left- and right-leaning users, as right-leaning users devote a substantially larger portion of their activity on Twitter to political communication. In fact, right-leaning users were almost twice as likely to create political content, with   
$22\%$ % STAT
of all tweets produced by right-leaning users containing one or more of the political hashtags under study, compared to only 
$12\%$ % STAT 
for left-leaning users. 

%%%%%% 
%
%  FIGURE: TWEET ACTIVITY
% 
%%%%%%%

\subsection{Partisan Self-Identification}
\label{sec:partisan_identification}
In addition to devoting a larger proportion of tweets to political content, right-leaning users are much more likely to use their $140$-character profile `biography' to explicitly self-identify their political alignment. 
%The biography is one of the few pieces of static content on a user's Twitter feed, accompanied only by a location field and optional website URL.  
A survey of the biographies of 400 random users from the set of individuals selected for qualitative content analysis (Section~\ref{sec:networks_of_political_communication}) reveals that 
$38.7\%$ % STAT
of right-leaning users included reference to their political alignment in this valuable space, as compared with only 
$24.6\%$ % STAT
of users in the left-leaning community. Taken together, this analysis demonstrates that right-leaning users are much more likely to use Twitter as an outlet for political communication, and are substantially more inclined to view the Twitter platform as an explicitly political space. 

\subsection{Resource Sharing}
\label{sec:hyperlinking}

One of the key functions of the Twitter platform is to serve as a medium for sharing information in the form of hyperlinks to external content~\cite{boyd08}.  Given the constraints of the 140-character format,  hyperlinking activity is especially important to the dissemination of detailed political information among members of a constituency. 

With respect to this aspect of online political engagement, too, we see that right-leaning users are more active then those individuals in the left-leaning community.  Among all tweets produced by users in the right-leaning community,  
$43.4\%$ contained a hyperlink, compared with $36.5\%$ % STAT
of all tweets from left-leaning users.  This trend is even more pronounced if we consider only resource sharing within the set of political tweets, with left-leaning users including a hyperlink in  $50.8\%$ of political tweets, as compared to right-leaning users, who include hyperlinks $62.5\%$ of the time. From these observations we conclude that right-leaning users are more inclined to treat Twitter as a platform for aggregating and sharing links to web-based resources, an activity crucial to the efficient spread of political information on the Twitter platform.

%%%%%%%%%%%%%%%%%%%%%%
%********* NETWORK TOPOLOGY ********%
%%%%%%%%%%%%%%%%%%%%%%

%%
\section{Connectivity: Global-level Political Activity}
\label{sec:network_structure}
Next, we turn our attention to structural differences in social interaction and communication networks of left- and right-leaning users. 

\subsection{Follower Network}
\label{sec:follower_networks}
\begin{figure}
\begin{center}
\includegraphics[width=\textwidth]{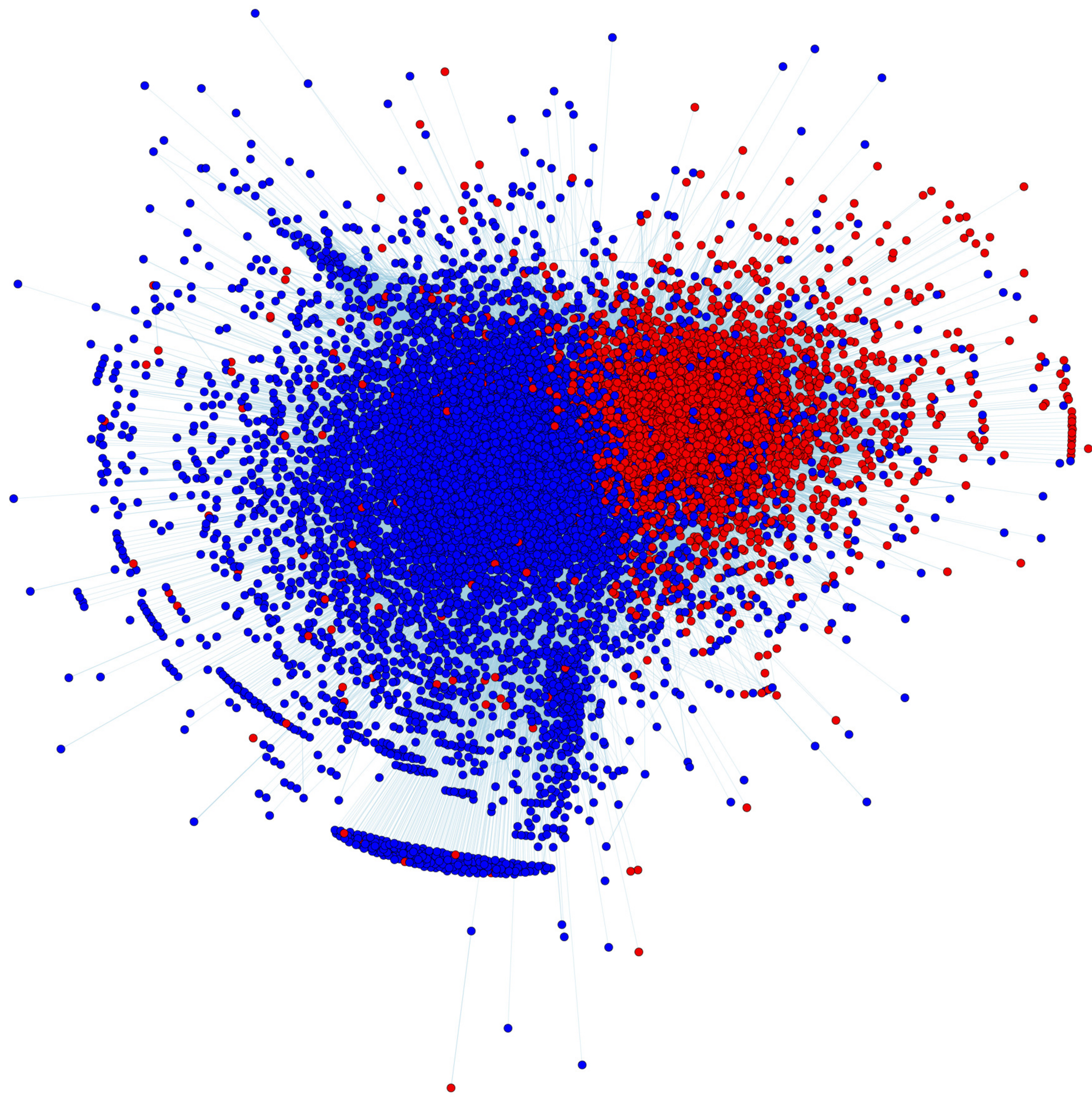}
\caption{\label{fig:follower_lr_layout}Force-directed layout of the follow relationships among politically-active Twitter users.  Nodes are colored according to political identity, Connections to users who did not engage political communication on Twitter are not included. }
\end{center}
\end{figure}

We begin with an analysis of the network defined by the follower/followee relationships shared among members of these two groups (Figure~\ref{fig:follower_lr_layout}). Encoding the fact that a user subscribes to the content produced by another, the follower network is best understood as describing the social substrate over which information is likely flow between political actors on Twitter.  Specifically, though not all connections in the follower network encode equally meaningful social relationships~\cite{huberman2009social}, content is broadcast equally along all edges  in this network. 

%% 
% STRUCTURE
%% 
We examine the differences in the follower subgraphs induced by considering only connections between users of the same political affiliation. For the purposes of this analysis, a directed edge is drawn from user $A$ to user $B$ if $A$ is a follower of $B$.  Basic statistics about these two subgraphs, including average degree, undirected clustering coefficient, and proportion of reciprocal links are presented in Table~\ref{tab:follower_lr_stats}. We see that along all dimensions, users in the right-leaning community are much more tightly interconnected, with a substantially higher average clustering coefficient and greater average degree.  Additionally, we observe a higher proportion of reciprocal links between right-leaning users, indicating the presence of stronger, mutually-affirmed interest among individuals in this community.  All of these factors indicate that right-leaning users are more tightly interconnected, resulting in a basic structural advantage with respect to the challenge of efficiently spreading political information on the Twitter platform.

%%%%%%
%
% TABLE: Follower Network Statistics
\begin{table*}
\caption{\label{tab:follower_lr_stats}Follower network statistics for the subgraphs induced by the set of edges among users of the same political affiliation.  Reciprocity is defined as $\frac{D_{R}}{D}$, where $D_{R}$ is the number of dyads with an edge in each direction and $D$ is the total number of dyads with at least one edge.  Follower data was only available for a subset of the study population, owing to private or deleted accounts.  }
\centerline{
	\begin{tabular}{lcccccc}
	\hline
	 & Community 			& Nodes 	& Edges 	& Avg. Degree & Clust. Coeff. 	& Reciprocity \\
	\hline
	\multirow{2}{*} 	& Left 	& $9,941$ & $803,329$  	& $80.80$   & $0.134$ & $42.8\%$ \\
				& Right 	& $6,426$ & $1,503,417$  & $233.95$   & $0.221$ & $64.8\%$ \\
		\hline
	\end{tabular}
}
\vspace{-1em}
\end{table*}
%
%%%%%%%

\begin{figure}
\begin{center}
\includegraphics[width=\textwidth,]{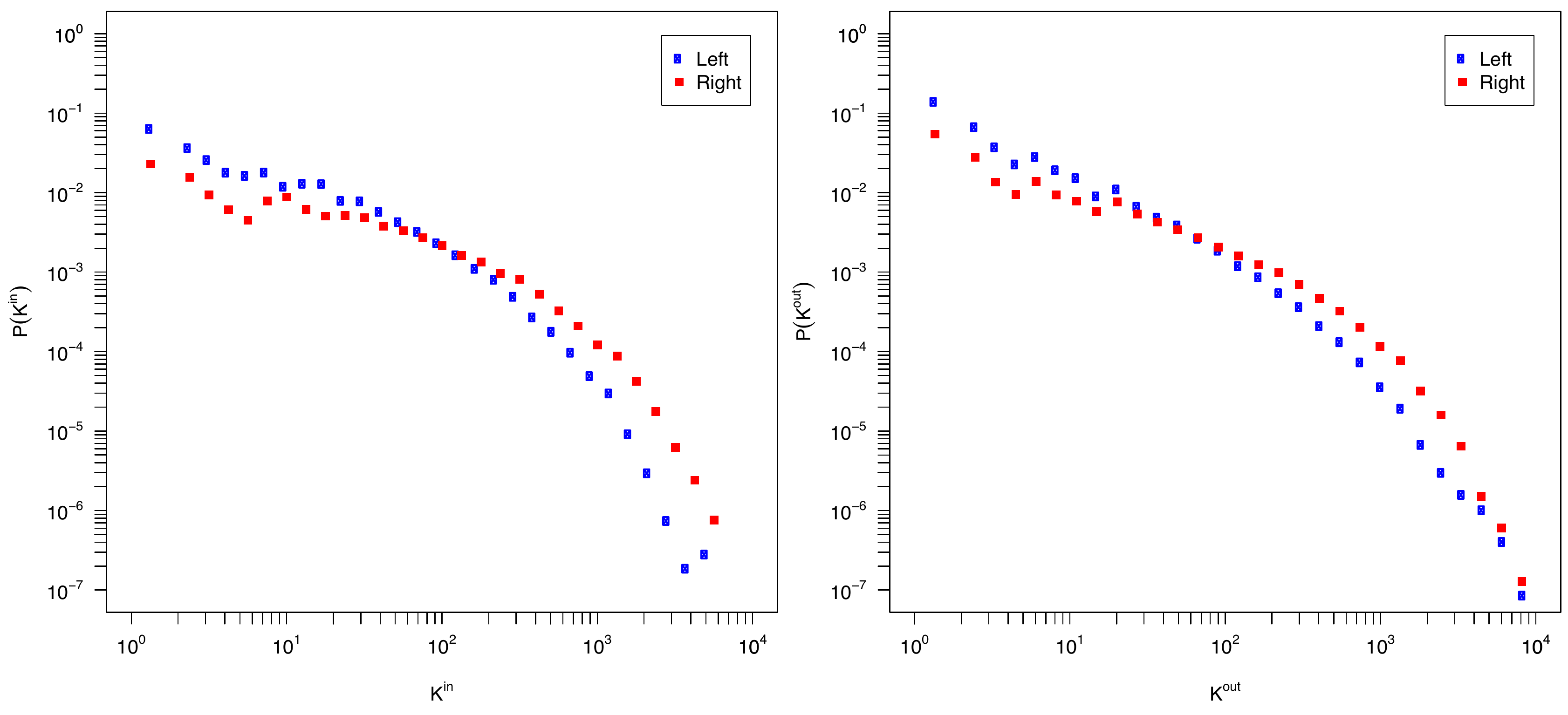}
\caption{\label{fig:follow_kin_lr_dist}Log binned in- and out-degree distributions of the internal follower network at left, and right, respectively.  As a result of considering only follower relationships among politically-active users we observe strong cutoffs in both distributions that make curve-fitting unreliable.  However, comparing the two distributions it's clear that the right-leaning community has a much greater proportion of users with many followers (Kolmogorov-Smirnov $p<10^{-3}$), despite being comprised of fewer users in total.  Understood as an information diffusion substrate, the proliferation of high-profile hubs gives a natural advantage to the right-leaning community. }
\end{center}
\end{figure}

% R&R 2.6
Using the Kolmogorov-Smirnov two-sample test to measure the degree of similarity between the in- and out-degree distribution for left- and right-leaning users we find a significant difference between the in-degree distributions of left- and right-leaning users, but only a marginal difference between the corresponding out-degree distributions ~(Figure~\ref{fig:follow_kin_lr_dist}). We interpret this to mean right-leaning users are more likely to have a large audience of followers who may potentially rebroadcast his or her call to action or piece of political information.   For example, left-leaning users are roughly twice as likely as right leaning users to have in-degree one, while users that are associated with the right are almost four times more likely to have in-degree $1000$ than users associated with the left.
%%%%

\begin{figure}
\begin{center}
\includegraphics[width=.65\textwidth,]{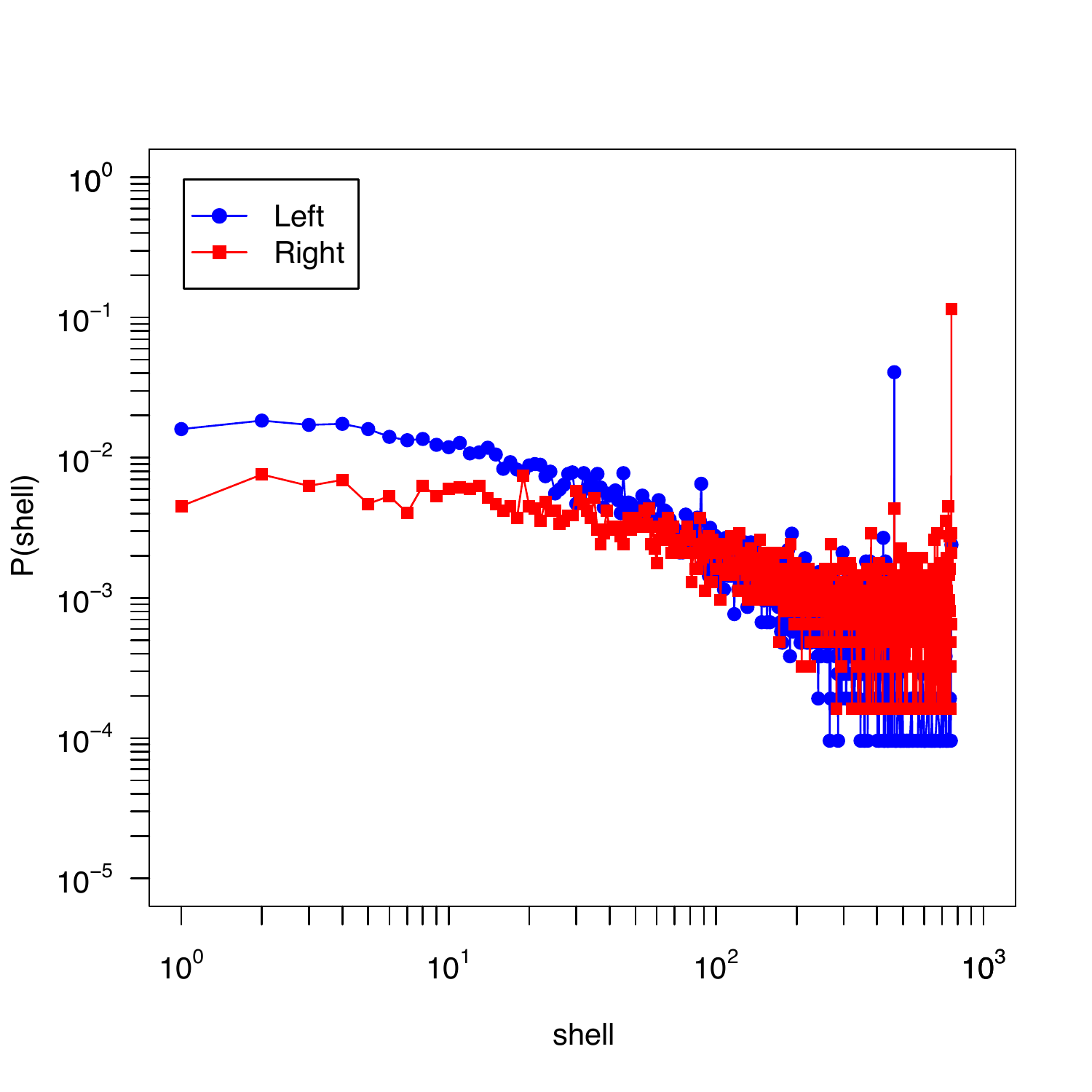}
\caption{\label{fig:cores-follow}Linearly binned core distribution of the internal follower network. The difference between these two distributions is highly significant (Kolmogorov-Smirnov $p<10^{-3}$).}
\end{center}
\end{figure}

 Additionally,  users in the left-leaning community are more likely to be only peripherally connected into the network, as evidenced by the distribution of the $k$-core shell indices of users in each community (Figure~\ref{fig:cores-follow}). For a given network, the $k$-core is the maximal subgraph whose nodes (as members of the subgraph) have at least degree $k$, or, in other words, have at least $k$ neighbors in the $k$-core itself. The shell index, $c$, of a node refers to the coreness ($k$) of the highest-order $k$-core of which the node is a member~\cite{barrat2008dynamical}.

These observations leads us to conclude that there are substantial structural differences in the fundamental patterns of social connectivity among politically left- and right-leaning Twitter users, a finding supported by the seminal work of Adamic \& Glance~\cite{Adamic:2005uq} on the connectivity patterns of high-profile partisan bloggers.  Specifically, the right-leaning community is much more densely interconnected, with more users tightly integrated into the right-leaning social network.  In contrast, the network of follower/followee relations among left-leaning users exhibits a much more decentralized, loosely-interconnected structure, with far fewer mutually-affirmed social connections.  

\subsection{Retweet Network}
\label{sec:retweet_network}

\begin{figure*}
\label{fig:network_layout}

\centerline{
\includegraphics[width=\textwidth,]{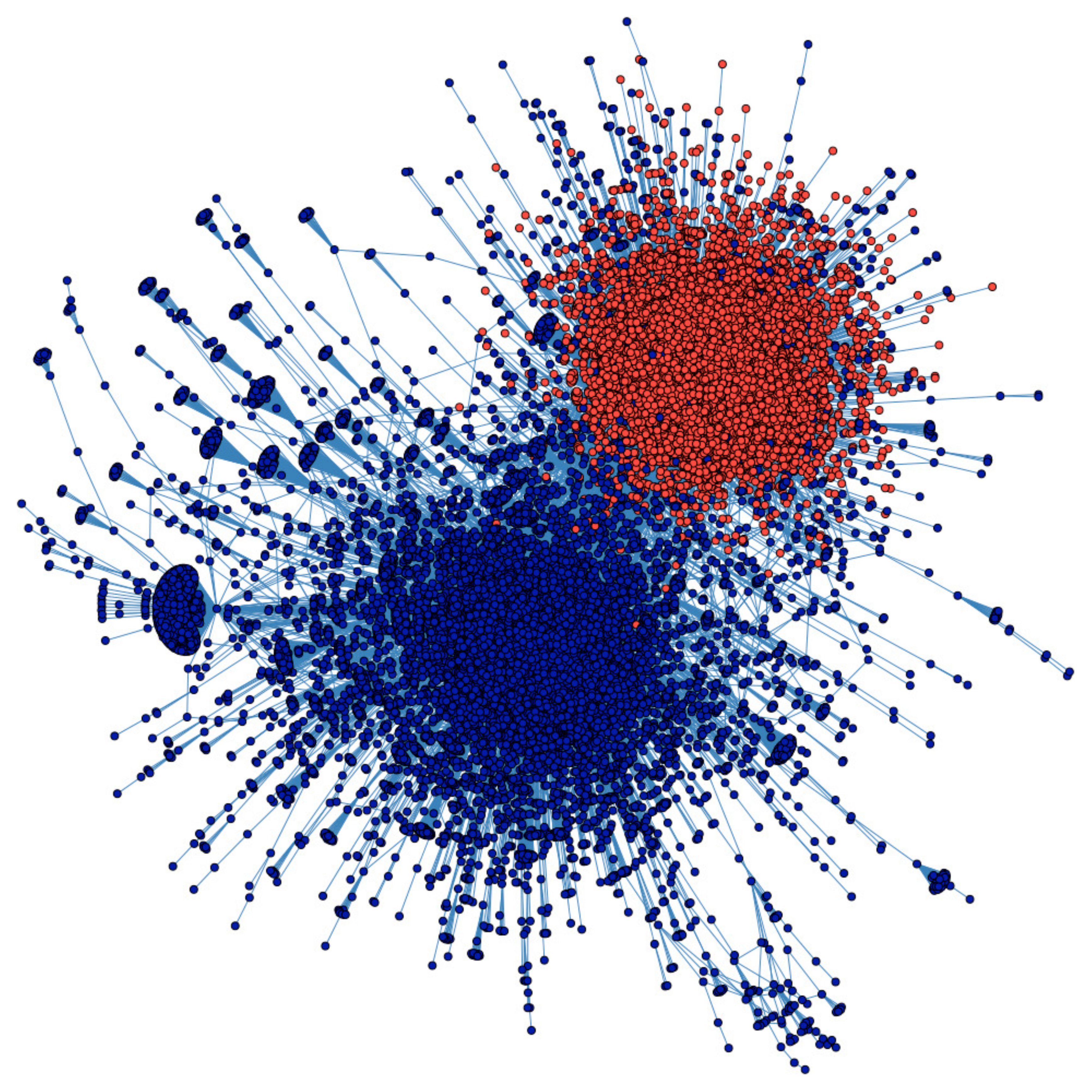}
}
\vspace{-1em}
\caption{\label{fig:retweet-network}The network of political retweets, laid out using a force-directed algorithm. Node colors reflect cluster assignments, which correspond to politically homogeneous communities of left- and right-leaning users with 87\% accuracy. (See \S~\ref{sec:networks_of_political_communication}).
}
\end{figure*}

Next we consider the structure of the network of political retweets in order to understand how information actually spreads on the social substrate characterized in Section~\ref{sec:follower_networks}.  While each link in the follower network represents a potential pathway along which information may flow, edges in the retweet network correspond  to real information propagation events.  Specifically, when user $A$ rebroadcasts a tweet produced by user $B$, she explicitly signifies receipt of the content in question, and thus we draw an edge from user $B$ to user $A$ indicating the direction of information flow.  Consequently, the structure of the retweet network reveals much about how information actually spreads within these two communities.  Visualized previously in Figure~\ref{fig:retweet-network}, basic statistics describing the networks induced by retweets containing at least one political hashtag between users of the same partisan affiliation are show in Table~\ref{tab:retweet_lr_stats}.

%%%%%%
%
% TABLE: Retweet Network Statistics
\begin{table*}
\caption{\label{tab:retweet_lr_stats}Retweet network statistics for the subgraphs induced by the set of edges among users of the same political affiliation.  }
\centerline{
	\begin{tabular}{lcccccc}
	\hline
	 & Community 			& Nodes 	& Edges 	& Avg. Degree & Clust. Coeff. 	& Reciprocity \\
	\hline
	\multirow{2}{*} 	
				& Left 	& $11,353$ 		& $32,772$	& $2.88$	& $0.032$	 & $13.5\%$    \\
				& Right 	&  $7,115$ 		& $39,713$ 	& $5.58$  	& $0.045$ & $12.1\%$     \\
		\hline
	\end{tabular}
}
\vspace{-1em}
\end{table*}
%
%%%%%%%

%%%%%% 
%
%  FIGURE: Retweet etwork Community Structure
% 
%
%%%%%%

\begin{figure}
\begin{center}
\includegraphics[width=\textwidth,]{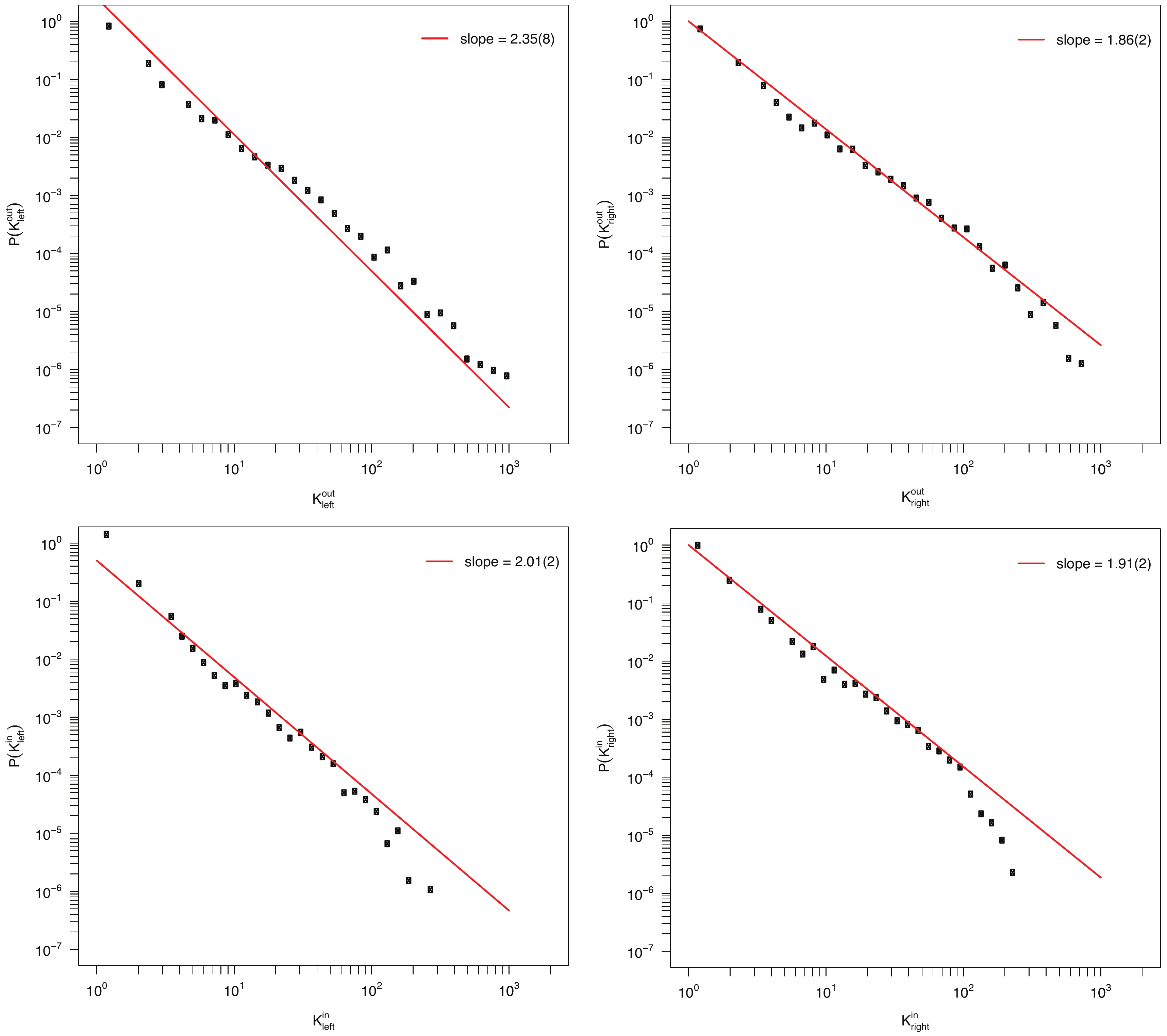}
\caption{\label{fig:deg_lr_retweet}Log binned in- and out-degree distributions for the left- and right-leaning retweet network communities. Slopes and standard errors were inferred using the maximum likelihood estimation method described by Clauset, Shalizi \& Newman~\cite{clauset2007power}. The rapid decay of the left-leaning degree distribution indicates that right-leaning users are retweeted by and retweet content from a larger number of users than those on the left.}
\end{center}
\end{figure}

In practice, the tightly-interconnected structure of the follower network confers communication advantages to the right-leaning community of users. Examining the in- and out-degree distributions for these two communities we find that though the power-law exponents are similar, the difference between them is statistically significant at the 95\% level (Figure~\ref{fig:deg_lr_retweet}). The faster decay in the degree distribution of the left-leaning community implies that right-leaning users are rebroadcast by and rebroadcast content from a larger number of individuals than users on the left. That right-leaning users pay attention to more information sources compared to left-leaning individuals is indicative of a higher degree of engagement with the Twitter platform itself.  Similarly, an individual wishing to rapidly reach a wide audience has a natural advantage given the structure of the right-leaning retweet network.

\begin{figure}
\begin{center}
\includegraphics[width=\textwidth,]{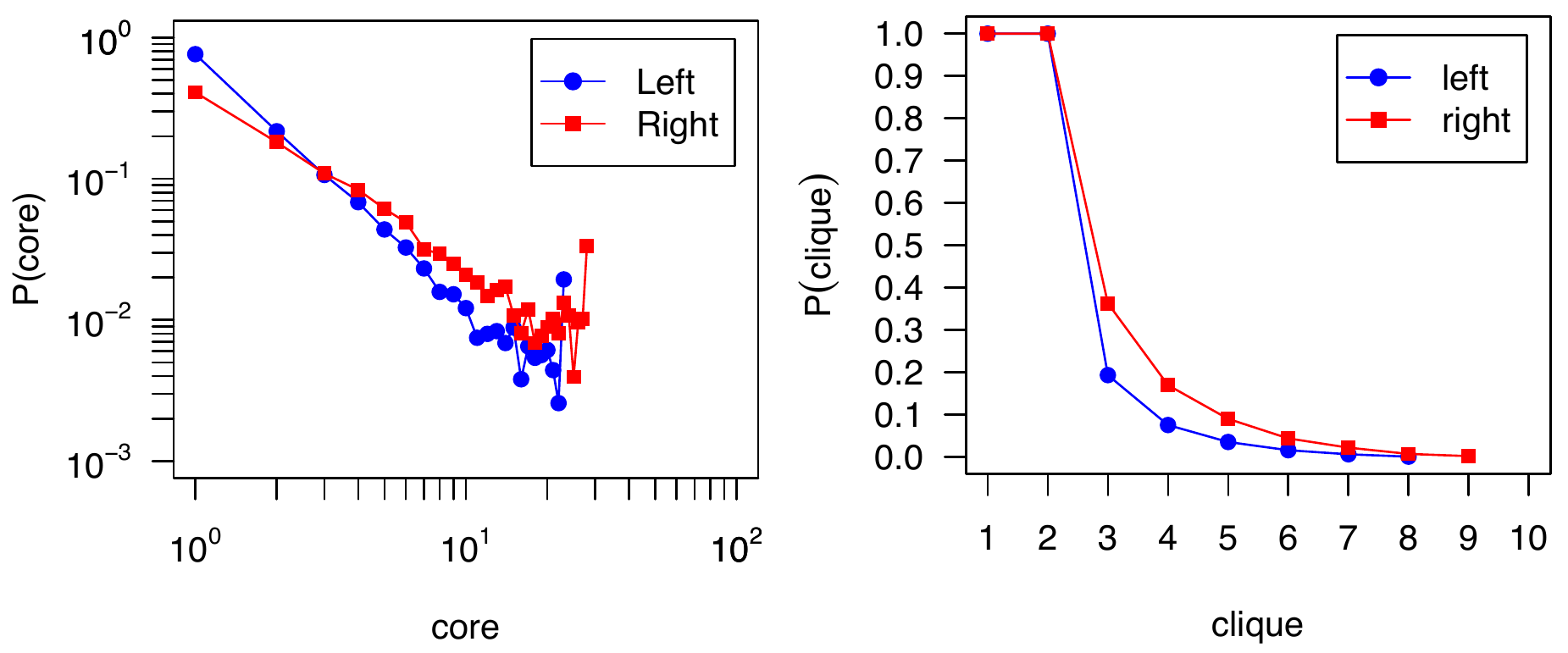}
\caption{\label{fig:cores_and_cliques}Proportion of users with a given $k$-core shell index (left) and membership in a k-clique (right) for the retweet network.}
\end{center}
\end{figure}

With respect to the number of users in high-order $k$-cores, too, we see that the right-leaning community enjoys structural advantages, with a greater proportion of highly active users connected to other highly active users (Figure~\ref{fig:cores_and_cliques}). This difference could lead to consequences in the spread of information through these networks. Work by Kitsak \emph{et al.} indicates that it is individuals with high shell index, rather than those who are most central or well connected, who are the most effective spreaders of information under a simple SIR-based information diffusion model~\cite{kitsak2010identifying}. Users on the right therefore, are more likely than those on the left to be wired into the political communication network in such a way that they are able to facilitate the broad and rapid dissemination of political information. 

We also find that a substantially higher proportion of right-leaning user participate in fully-connected subgraphs of size $k$, known as k-cliques.  This result is especially important in the context of the complex contagion hypothesis, which posits that repeated exposures to controversial behaviors are essential to the adoption of these behaviors. Work by Romero, Meeder and Kleinberg focused specifically on online social networks indicates that this effect is particularly pronounced for political discourse on Twitter~\cite{romero2011differences}. With fewer users in high-order $k$-cores, individuals in the left-leaning community will be less likely to encounter multiple users discussing the same partisan talking points or calls to action, exactly the kind of contentious content whose propagation is most likely to benefit from repeated exposure.  

\subsection{Mention Network}
\label{sec:mention_network}
% R&R: 2.1
Mentions are most strongly associated with direct, conversational engagement when the target username appears at the beginning of a tweet, as opposed to appearing in the body text. Among the mentions in our sample, the overwhelming majority (94.5\%) take this form, providing strong evidence that connectivity among and between users in these two groups represents actual political discourse rather than simply third-person references.  In Table~\ref{tab:mention_lr_stats} we report descriptive statistics on the topology of the left- and right-leaning mention networks, where an edge from $A$ to $B$ is drawn between two users of the same political affiliation if $A$ mentions $B$ in a tweet containing at least one political hashtag. Though the two networks exhibit very similar degree distributions, one important distinction is the fact that a greater proportion of mention relationships in the right-leaning community are reciprocal. Compared to the number of reciprocal mentions observed in degree-preserving reshufflings of the left- and right-leaning mention networks, the right-leaning community exhibits $7.5$ times as many reciprocal mention interactions than is expected by chance alone, compared to a $5.6$ times as many reciprocal links in the left-leaning community. Reciprocal interactions suggest the presence of more meaningful social connections, manifest in conversational dialogue, rather than, for example, unidirectional commentary on the content of another user's tweets~\cite{huberman2009social}.  Here too, we find that users on the political right are more  engaged with one another on Twitter, indicating that they are likely to benefit from a richer dialogue and hence more opportunities for frame-making and consensus building with respect to political topics. 

%%%%%%
%
% TABLE: Mention Network Statistics
\begin{table*}
\caption{\label{tab:mention_lr_stats}Mention network statistics for the subgraphs induced by the set of edges among users of the same political affiliation.}
\centerline{
	\begin{tabular}{lcccccc}
	\hline
	 & Community 			& Nodes 		& Edges 		& Avg. Degree	 		& Clust. Coeff. 	& Reciprocity \\
	\hline
	\multirow{2}{*} 	
				& Left 	& $11,353$ 	& $50,273$	& $4.42$  	& $0.053$		& $20.8\%$     \\
				& Right 	&  $7,115$ 	& $64,993$ 	& $9.13$   & $0.078$ 	& $24.5\%$     \\
		\hline
	\end{tabular}
}
\vspace{-1em}
\end{table*}

%Given the result of the $2010$ midterm elections, our data suggest that it is the character and vigor of engagement, rather than the raw number of people of a specific ideology involved in the online political discourse, that is correlated with positive electoral outcomes.

%% SECTION: Cartogram
\section{Political Geography}

In addition to characterizing differences in behavior and connectivity, we can also examine the geographic distribution of individuals in these two communities.  Here we present a cartogram in which the color of each state has been scaled to correspond to the degree to which, in that state, the observed number of tweets originating from the left-leaning community exceeds what we should expect by chance alone.  

Because fewer than one percent of Twitter users provide precise geolocation data, we instead rely on the self-declared `location' field of each user's profile to enable geographic analysis of data at the scale of this study. As a free-text field, users are able to enter in arbitrary data, and non-location responses such as `the moon' do appear in the results.  Complicating this analysis further, some users do not report any location data, though we do not report a partisan bias in terms of non-entries. Despite these caveats, a large number of users do report actual locations, and using the Yahoo Maps Web Service API\footnote{\emph{http://developer.yahoo.com/maps/rest/V1/geocode.html}}, we are able to make a best-guess estimate about the state with which a user most strongly identifies.  

\begin{figure}
\begin{center}
\centerline{
	\includegraphics[width=\textwidth,]{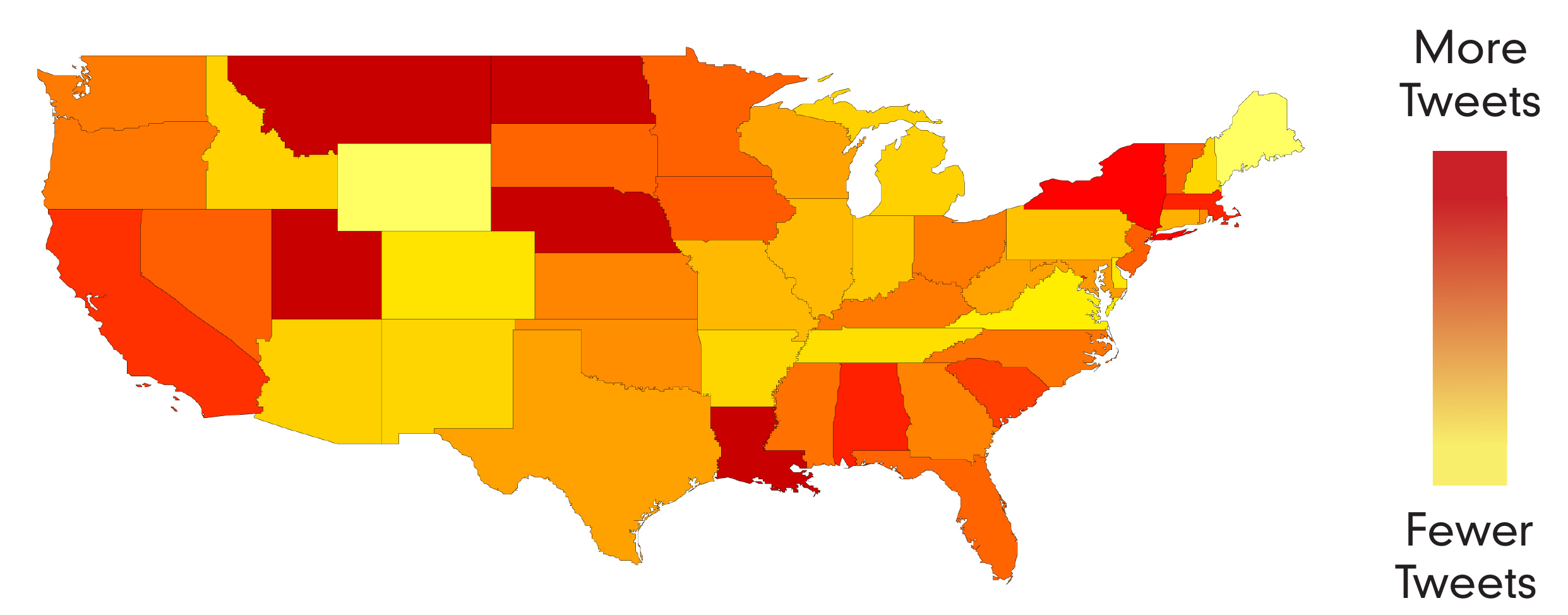}
}
\vspace{-1em}
\caption{Deviation in volume of left-leaning political communication compared to expected baseline. Each state is filled with a color corresponding to the extent to which the observed number of tweets is above or below what should be expected in the case where each state has traffic volume proportional to that observed across all Twitter traffic.}
\label{fig:ab_cluster_origins}
\end{center}
\end{figure}

Thus, for each state in which we observe $N$ total tweets, and the relative proportion of tweets originating from left-leaning users ($P_{l}$), we can treat the arrival of partisan tweets as a Bernoulli process, and compute the number of tweets we should expect to see from left-leaning users as $NP_{l}$.  Likewise, we can compute the extent to which the observed number of tweets associated with left-leaning users ($T_{l}$) is above or below the expected number, measured in terms of standard deviations, as $\frac{T_{l}-NP_{l}}{\sqrt{ NP_{l} \cdot (1-P_{l}) }}$.  Figure~\ref{fig:ab_cluster_origins} uses color to encode these deviations for each state, with states in which the volume of activity far exceeds what should be expected by chance shown in deep red, and those in which the observed volume is far below what should be expected by chance shown in light yellow.

% R&R 1.2.1
%%%%%% 
%
%  FIGURE: TWEET ACTIVITY
% 
%%%%%%%

Initial inspection of this figure reveals that the geographic distribution of individuals from the left-leaning network community corresponds strongly to the traditional political geography of the United States. We see that left-leaning individuals feature prominently on the coasts and North East, and tend to be underrepresented in the midwest and plains states. 

% R&R 2.4.3
Looking more closely, however, we find that there are some places in which the partisan makeup of tweets is quite different from what might be hypothesized intuitively.  For example, Utah, a traditionally conservative state which at the time of this writing had two Republican senators, exhibits a dramatically higher volume of left-leaning content than should be expected by chance alone. One possible explanation for this observation could be that individuals in some states with a ideologically homogeneous population turn to social media as an outlet for political expression.  While this is but one possible explanation among many, and a more rigorous analysis is required to support any definitive claim, this example illustrates the ways in which novel hypotheses can derive from data-driven analyses of political and sociological phenomena.

\section{Conclusion}
In this study we have described a series of techniques and analyses that indicate a shifting landscape with respect to partisan asymmetries in online political engagement. We find that, in contrast to what might be expected given the online political dynamics of the $2008$ campaign, right-leaning Twitter users exhibit greater levels of political activity, tighter social bonds, and a communication network topology that facilitates the rapid and broad dissemination of political information.  

In terms of individual behavior, politically right-leaning Twitter users not only produce more  political content and devote a greater proportion of their time to political discourse, but are also more likely to view the Twitter platform as an explicitly political space and identify their political leanings in their profiles.  With respect to social interactions, the right-leaning community exhibits a higher proportion of reciprocal social and mention relationships, are more likely to rebroadcast content from a large number of sources, and are more likely to be members of high-order retweet network $k$-cores and $k$-cliques. Such structural features  are directly associated with the efficient spreading of information and adoption of political behavior.  Taken together, these features are indicative of a highly-active, densely-interconnected constituency of right-leaning users using this important social media platform to further their political views.

This study is characteristic of an emerging mode of inquiry in the political and social sciences, whereby large-scale behavioral data are aggregated and analyzed to shed quantitative light on questions whose scale was previously considered outside the realm of tractable analysis~\cite{lazer2009life}.  Using structural features of a digital communication network one can make high-fidelity inferences about the political identities of thousands of individuals. Such data provide a deeper understanding of the changing landscape of American online political activity. Looking forward, techniques such as these are likely to become increasingly important as the political and social sciences rely in greater measure on large-scale digital trace data describing human opinion and behavior.

\bigskip

%%%%%%%%%%%%%%%%%%%%%%%%%%%%%%%%
\section*{Author's contributions}
MDC and BG collected the data and performed the analysis. MDC, BG, AF and FM conceived the experiments and wrote the manuscript.

\bibliographystyle{abbrvnat}
\bibliography{references}
\end{document}